# Application des techniques d'enchères dans les réseaux de radio cognitive

**Badr Benmammar**

*Rapport de recherche*
*Laboratoire de Télécommunications Tlemcen*
*Université Abou Bekr Belkaid , Tlemcen, Algérie*
*BP 230 pôle Chetouane, 13000 Tlemcen*
*badr.benmammar@gmail.com*

RESUME. La prolifération rapide de standards et services de radiocommunication ces dernières années provoquent le problème de la pénurie du spectre. Dans ce contexte, l'objectif principal de la Radio Cognitive (RC) est de faciliter l'accès au spectre radio. Notre contribution dans le cadre de ce papier est l'utilisation des enchères pour résoudre le problème de l'encombrement du spectre dans le cadre de la RC. Pour cela, nous avons combiné la théorie des enchères avec les systèmes multi agents. Notre approche a prouvé qu'il est préférable d'utiliser les enchères à Enveloppe Scellée avec programmation dynamique car cette méthode a beaucoup d'avantages par rapport aux autres méthodes.

ABSTRACT. The rapid proliferation of standards and radio services in recent years caused the problem of spectrum scarcity. The main objective of Cognitive Radio (CR) is to facilitate access to radio spectrum. Our contribution in this paper is the use of auctions to solve the problem of spectrum congestion in the context of CR, for that, we will combine the theory of auctions with multi-agent systems. Our approach has shown that it is preferable to use the Sealed-bid Auction with dynamic programming because this method has many advantages over other methods.

# 1. Introduction

Avec la prolifération des différentes technologies de réseau sans fil tels que les réseaux cellulaires, réseaux locaux sans fil, les réseaux métropolitains sans fil, etc. la demande de spectre radioélectrique est en augmentation considérable. Actuellement, le spectre est réglementé par un organisme gouvernemental comme la Federal Communications Commission (FCC), et il alloue le spectre en attribuant des licences exclusives pour les utilisateurs d'exploiter leurs réseaux dans différentes régions géographiques.

Il y a une croyance largement répandue que le spectre radioélectrique est de plus en plus encombré. Toutefois, les mesures de spectre indiquent que le spectre attribué est sous-utilisé, c'est à dire, à n'importe quel temps et endroit données, une grande partie du spectre est non utilisée. Les réseaux de radio cognitive (RC) apparaissent comme une solution prometteuse à ce problème.

L'idée de la RC consiste à partager le spectre entre un utilisateur dit secondaire qui pourra à tout moment accéder à des bandes de fréquence qu'il trouve libres, et un utilisateur dit primaire qui possède une licence sur cette bande.

Il faut noter également que l'évolution technologique de ces dernières décennies a entraîné une croissance considérable de nouveaux services de télécommunications utilisant la ressource rare que constitue le spectre radioélectrique. On prévoit que la demande de fréquences continuera d'augmenter avec le développement de nouvelles techniques de télécommunications ce qui augmente l'allocation du spectre statique qui est un problème majeur dans les réseaux sans fil. Généralement, ces allocations conduisent à une utilisation inefficace du spectre donc le problème de l'encombrement de celui-ci.

Dans le cadre de ce travail, nous avons utilisé la théorie des enchères qui est une technique d'accès dynamique au spectre pour résoudre le problème d'encombrement. Cette méthode est connue par sa simplicité dans le but de faciliter l'allocation des ressources rares. Donc, dans le cadre de ce papier, nous avons mis en place plusieurs méthodes tels que une variante de FIFO (FIFO' : FIFO mais sans blocage des demandes insatisfaites), l'enchère à enveloppe scellée avec et sans programmation dynamique. Les trois méthodes ont été implémentées à l'aide de l'outil JADE (Java Agent DEvelopment Framework). Une comparaison entre les trois méthodes a été également réalisée afin de tirer des conclusions dans ce contexte. Les résultats obtenus nous ont permis de confirmer que l'utilisation des enchères avec programmation dynamique a beaucoup d'avantages en termes de nombre de SUs satisfaits, gain obtenu par le PU, temps de traitement côté PU et finalement temps de réponse coté SU.

Ce papier est organisé comme suit :

Dans la section 2, nous introduisons le concept de la radio cognitive tout en présentant sa définition et ses principes les plus importantes, ensuite les différentes phases de cycle de cognition et les fonctions la radio cognitive seront détaillées.

Enfin, une brève description sera donnée sur les réseaux de la radio cognitive pour comprendre le fonctionnement exact de ce paradigme.

La troisième section présente une brève description sur les diverses méthodes d'accès dynamique au spectre tels que la théorie des jeux, les chaînes de Markov et les systèmes multi agent. Tandis que, concernant notre sujet principal, nous nous concentrons sur les approches d'accès au spectre basées sur la théorie des enchères.

La quatrième section décrit les simulations que nous avons effectuées pour valider notre approche. Pour mettre en pratique ceci, nous avons mis en place un scénario qui décrit notre approche auquel nous sommes confrontées. Une fois que les résultats sont obtenus, une comparaison est effectuée pour montrer que l'utilisation des enchères avec programmation dynamique est la solution la plus efficace pour améliorer la gestion et l'utilisation du spectre radio.

**2. Radio cognitive**

La radio cognitive est un nouveau paradigme de la conception de systèmes de communication sans fil qui vise à améliorer l'utilisation de la fréquence radio (RF) du spectre. La motivation derrière la radio cognitive est la rareté du spectre des fréquences disponibles, la demande croissante, provoquée par les applications sans fil émergents pour les utilisateurs mobiles.

Le principal facteur causé par la forte demande est l'encombrement de spectre qui conduit à une utilisation inefficace de celui-ci. Pour résoudre ce problème il faut une bonne gestion du spectre. C'est dans ce cadre que des études sont menées dans le domaine de la radio cognitive par des chercheurs donc il est clair que les réseaux de radio cognitive constituent un domaine de recherche très prometteur.

*2.1. Définition et principe*

L'idée de la radio cognitive a été présentée officiellement par Joseph Mitola III à un séminaire à KTH, l'Institut royal de technologie, en 1998, publié plus tard dans un article de Mitola et Gerald Q. Maguire, Jr en 1999.

Mitola combine son expérience de la radio logicielle ainsi que sa passion pour l'apprentissage automatique et l'intelligence artificielle pour mettre en place la technologie de la radio cognitive. D'après lui : « Une radio cognitive peut connaître, percevoir et apprendre de son environnement puis agir pour simplifier la vie de l'utilisateur. »

Le terme radio cognitive (RC) est utilisé pour décrire un système ayant la capacité de détecter et de reconnaître son cadre d'utilisation, ceci afin de lui permettre d'ajuster ses paramètres de fonctionnement radio de façon dynamique et autonome et d'apprendre des résultats de ses actions et de son cadre environnemental d'exploitation. (cf Benmammar et Amraoui., 2013)

La radio cognitive est une technologie qui fait appel à l'intelligence des réseaux et des terminaux pour :

- détecter les besoins de communications des utilisateurs en fonction de l'utilisation ;

- fournir des ressources radio et les services sans fil les plus appropriés à ces besoins.

Une radio cognitive doit être capable de réaliser trois tâches essentielles :

**- « Aware »** : c'est la capacité à prendre conscience de son environnement. Un terminal radio cognitif associera donc environnement spatial, spectral et comportement des usagers, pour une meilleure prise en conscience du provisionnement en ressources et un meilleur service.

**- « Adaptation »** : c'est la capacité à s'adapter soit à l'environnement (spectral ou technologique), soit à l'utilisateur (besoins et sécurité).

L'adaptation à l'environnement spectrale est la capacité à choisir les meilleures bandes de fréquences et ainsi optimiser l'utilisation du spectre. Cela revient à :

- connaître l'occupation des bandes de fréquences en temps réel ;
- adapter les puissances émises / marges en réception / formes d'ondes ;
- adapter le débit en temps réel en fonction de la place disponible ;
- prendre en compte un partage temporel.

**- « Cognition »** : La cognition regroupe les divers processus mentaux allant de l'analyse perceptive de l'environnement à la commande motrice (en passant par la mémorisation, le raisonnement, les émotions, le langage…). Cette définition dépasse donc le seul cadre de la cognition humaine ou animale. La radio cognitive est donc un système qui peut percevoir son environnement, l'analyser, le mémoriser et agir en conséquence. (cf Chehata., 2011)

La radio cognitive offre également une solution équilibrée au problème de l'encombrement du spectre en accordant d'abord l'usage prioritaire au propriétaire du spectre, puis en permettant à d'autres de se servir des portions inutilisées du spectre. (cf Benmammar et Amraoui., 2013)

*2.2. Cycle de cognition*

Une radio cognitive est un système radio qui opère selon un cycle précise appelé «cycle cognitif». Mitola définit le « cycle cognitif » en six étapes selon le schéma générique présenté dans la Figure 1 :

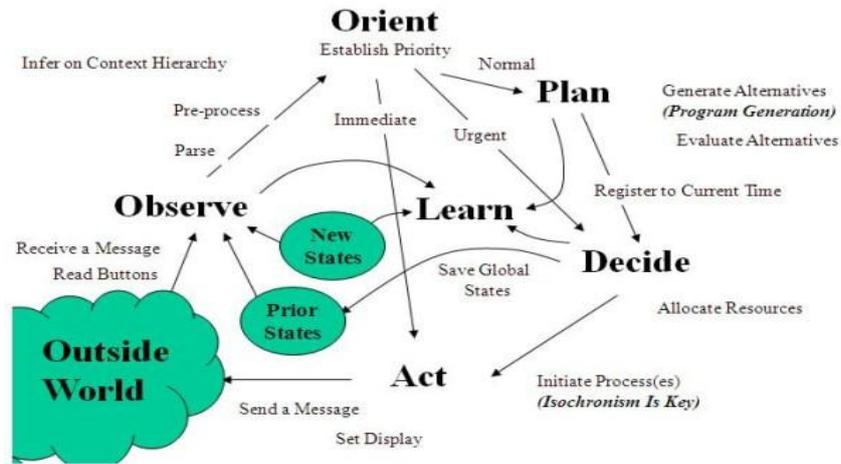

*Figure 1. Cycle de cognition version Mitola.*

Ce « cycle cognitif » se réalise en parallèle avec la fonction principale du système radio, c'est à dire la communication. (cf Colson et Kountouris, 2006). Une radio cognitive fonctionne ainsi :

**- Phase d'observation** : La RC observe son environnement par l'analyse du flux de stimuli entrant. Dans la phase d'observation, la RC associe l'emplacement, la température, le niveau de lumière des capteurs, et ainsi de suite pour en déduire le contexte de communication. Cette phase lie ces stimuli à des expériences antérieures pour discerner les modèles au fil du temps. La radio cognitive rassemble les expériences en se souvenant de tout.

**- Phase d'orientation :** La phase d'orientation détermine l'importance d'une observation en liant à celle-ci une série connue de stimuli. Cette phase fonctionne à l'intérieur des structures de données qui sont analogues à la mémoire à court terme (STM), que les gens emploient pour s'engager dans un dialogue sans forcément se souvenir de tout à la même mesure que dans la mémoire à long terme (LTM). Le milieu naturel fournit la redondance nécessaire pour lancer le transfert de la STM à la LTM. La correspondance entre les stimuli courants et les expériences stockées se fait par reconnaissance des stimuli ou par reliure.

**- Phase de planification** : La plupart des stimuli sont traités avec délibérative plutôt qu'avec réactivité. Un message entrant du réseau serait normalement traité par la génération d'un plan (dans la phase de plan, la voie normale). Le plan devrait également inclure la phase de raisonnement dans le temps. Généralement, les réponses réactives sont préprogrammées ou apprises en étant dit, tandis que d'autres réactions de délibération sont prévues.

**- Phase d'action** : Cette phase lance les processus sélectionnés qui utilisent les effecteurs sélectionnés qui accèdent au monde extérieur ou aux états internes de la radio cognitive. L'accès au monde extérieur consiste principalement à composer des messages qui doivent être envoyés dans l'environnement en audio ou exprimés dans différents langages appropriés.

**- Phase de décision** : La phase de décision sélectionne un plan parmi les plans candidats. La radio peut alerter l'utilisateur d'un message entrant ou reporter l'interruption à plus tard en fonction des niveaux de QoI (Quality of Information) statués dans cette phase, en fonction de cette information, la radio décide si une bande du spectre est libre ou si elle est occupée ; si elle est occupée, une autre bande libre est choisie pour la poursuite de la communication.

**- Phase d'apprentissage** : L'apprentissage dépend de la perception, des observations, des décisions et des actions. L'apprentissage initial est réalisé à travers la phase d'observation dans laquelle toutes les perceptions sensorielles sont continuellement comparées à l'ensemble de l'expérience antérieure pour continuellement compter les événements et se souvenir du temps écoulé depuis le dernier événement. (cf Benmammar et Amraoui., 2013)

*2.3. Fonctions de la radio cognitive*

Les principales fonctions de la radio cognitive sont les suivants : (cf Amraoui et al., 2012)

**- *Détection de spectre (Spectrum sensing)* :** Détection de spectre inutilisé et de le partageant, sans interférences nuisibles à d'autres utilisateurs ; une condition importante du réseau de radio cognitive pour capter le spectre vide. Détecter les utilisateurs primaires est le moyen le plus efficace pour détecter le spectre vide. Les techniques de détection de spectre peuvent être regroupées en trois catégories :

  **- Détection d'émetteur :** Les radios cognitives doivent avoir la capacité de déterminer si un signal d'un émetteur primaire est présent localement dans un certain spectre. Il existe plusieurs approches proposées à la détection d'émetteur :

   - détection d'un filtre correspondant (Matched filter detection).
   - détection d'énergie (Energy detection).
   - détection de fonctionnalité cyclostationnaire (Cyclostationary-feature detection).

  **- Détection coopérative** : Fait référence aux méthodes de détection du spectre où les informations provenant de plusieurs utilisateurs de radio cognitive sont incorporées pour la détection de l'utilisateur primaire.

  **- Détection d'interférences.**

**- *La gestion du spectre (Spectrum management)* :** Capturer le meilleur spectre disponible pour répondre aux besoins de communication des utilisateurs tout en ne créant pas l'interférence anormale aux autres utilisateurs (primaires). Les radios cognitives devrait décider de la meilleure bande de fréquences (de toutes les bandes disponibles) pour répondre à la qualité des services exigences et par conséquent , les fonctions de gestion du spectre sont requises pour les radios cognitives. Le fonctions de gestion du spectre sont classées comme : (cf Amraoui et al., 2012)

- **Analyse du spectre** : elle permet de caractériser les différentes bandes spectrales en termes de fréquence d'opération, de débit, de temps et de l'activité de l'utilisateur primaire. Cette caractérisation sert à répondre aux exigences de l'utilisateur. Des paramètres supplémentaires viennent compléter cette caractérisation, à savoir, le niveau d'interférence, le taux d'erreur du canal, les évanouissements, le délai et le temps d'occupation de la bande spectrale par un utilisateur. (cf Chehata., 2011)

- **Décision sur le spectre** : après que toutes les bandes spectrales aient été catégorisées et classifiées, on applique un ensemble de règles décisionnelles pour obtenir la ou les bandes spectrales les plus appropriées à la transmission en cours, en tenant compte des exigences de l'utilisateur. (cf Chehata., 2011)

- *Modèle de décision :* un modèle de décision est nécessaire pour l'accès au spectre. La complexité de ce modèle dépend des paramètres considérés lors de l'analyse du spectre.

- *Compétition / coopération dans un environnement multi utilisateurs :* Lorsque plusieurs utilisateurs (à la fois primaires et secondaires) sont dans le système, leur préférence va influer sur la décision du spectre d'accès. Ces utilisateurs peuvent être coopératifs ou non coopératifs dans l'accès au spectre.

Dans un environnement non-coopératif, chaque utilisateur a son propre objectif, tandis que dans un environnement coopératif, tous les utilisateurs peuvent collaborer pour atteindre un seul objectif. Dans un environnement coopératif, les radios cognitives coopèrent les unes avec les autres pour prendre une décision pour accéder au spectre et de maximiser une fonction objectif commune en tenant compte des contraintes. Dans un tel scénario, un contrôleur central peut coordonner le spectre de gestion. (cf Benmammar et Amraoui., 2013)

- *La mobilité du spectre (Spectrum mobility) :* processus par lequel un utilisateur de radio cognitive modifie sa fréquence de fonctionnement. Les réseaux de radio cognitive visent à utiliser les fréquences de manière dynamique, en permettant aux terminaux radio fonctionnant dans la meilleure bande de fréquence disponible, maintenant des exigences de communication transparente lors des transitions à un meilleur spectre. (cf Amraoui et al., 2012)

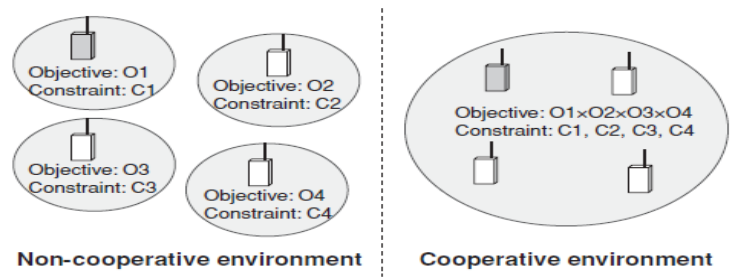

*Figure 2. Accès au spectre coopératif et non-coopératif. (Ekram et al., 2009)*

### *2.4. Architecture d'un réseau de radio cognitive*

Les réseaux de radio cognitive (ou smart) sont une approche innovatrice à l'ingénierie sans fil dans lequel les radios sont conçus avec un niveau sans précédent de l'intelligence et l'agilité. Cette technologie de pointe permet aux appareils de radio d'utilisation du spectre dans des voies entièrement nouvelles et sophistiquées.

Le réseau radio cognitive est un système de communication multi-utilisateurs complexes sans fil capable d'un comportement émergent. Elle incarne les fonctions suivantes (cf Simon., 2008):

- percevoir l'environnement radio (c'est à dire, à l'extérieur du monde) en habilitant le récepteur de chaque utilisateur pour détecter l'environnement sur une base continue dans le temps ;

- apprendre de l'environnement et d'adapter les performances de chaque émetteur/récepteur (émetteur-récepteur) aux variations statistiques dans les stimuli RF entrants ;

- faciliter la communication entre plusieurs utilisateurs grâce à la coopération d'une manière auto-organisés ;

- contrôler les processus de communication entre les utilisateurs concurrents grâce à une répartition judicieuse des ressources disponibles ;

- créer l'expérience de l'intention et la conscience de soi.

Une description détaillée de l'architecture des RRC est primordiale pour développer des protocoles efficaces de communication. Dans cette section, nous décrirons l'architecture des RRC qui peut être décomposé en deux groupes : réseau primaire et réseau secondaire.

**- Le réseau primaire** : est le réseau doté d'une licence pour utiliser certaines bandes spectrales. Le réseau primaire a acquis ce droit à travers l'achat de licences des agences gouvernementales. Il est composé des éléments suivants (cf Ahmed Ben, 2011) :

- **Un utilisateur primaire** : (dit utilisateur licencié) est un utilisateur qui détient une licence pour opérer sur des bandes spectrales qui lui sont réservées. L'accès est contrôlé uniquement par ses stations de base et ne doit pas subir d'interférence extérieure nuisible. Les PUs ne doivent subir aucune modification pour permettre la coexistence avec les utilisateurs ou réseaux de radios cognitives ou leurs stations de base.

- **Une station de base primaire** : (dite station de base licenciée) La station de base des PU est une structure fixe du réseau primaire qui possède une licence pour opérer sur la bande spectrale. Ces stations sont conçues de telle sorte à ne pas partager le spectre avec aucune entité extérieure au système, en l'occurrence, les utilisateurs à radio cognitive. Cependant, il peut exister des stations de bases licenciées qui reconnaissent les protocoles des utilisateurs à radio cognitive.

- **Le réseau secondaire** : Le réseau secondaire, appelé aussi réseau de radios cognitives ou réseau non-licencié, ne possède pas de licence pour opérer sur la bande spectrale. D'où le besoin d'un ensemble de fonctionnalités additionnelles pour pouvoir partager les bandes spectrales licenciées d'une manière opportuniste. Les réseaux secondaires sont déployés en mode infrastructure ou en mode ad-hoc. Ils se disposent de quatre composants (cf Ahmed Ben, 2011) :

- **Un utilisateur à radio cognitive :** (dit utilisateur non licencié ou utilisateur secondaire) n'a pas de licence pour transmettre sur la bande spectrale. Cependant, grâce aux fonctionnalités additionnelles dont ils disposent, ces utilisateurs pourront partager la bande spectrale avec les utilisateurs primaires à condition de ne pas nuire leurs transmissions ou bien profiter de leur absence pour transmettre.

- **Une station de base secondaire :** (dite station de base non licenciée) est une infrastructure fixe avec des capacités cognitives. L'utilisateur se connecte à la station de base secondaire pour accéder à d'autres réseaux ou services.

- **Un serveur spectral :** (en anglais Spectrum server) est une entité du RRC qui sert à partager les ressources spectrales entre différents utilisateur dans le même réseau. Ce serveur est connecté aux réseaux secondaires et agit comme un gestionnaire d'information spectrale. Dans un RRC, un utilisateur peut être élu par les autres utilisateurs pour remplir les tâches du serveur spectral. L'élection de ce serveur dépend essentiellement de l'emplacement de ce dernier et de ses capacités à réaliser une détection spectrale quasi parfaite.

- **Un courtier spectral :** (en anglais Spectrum broker) est une entité du RRC qui partage les ressources spectrales entre différents RRC. Ce serveur est connecté à plusieurs RRC et agit comme un gestionnaire d'information spectrale.

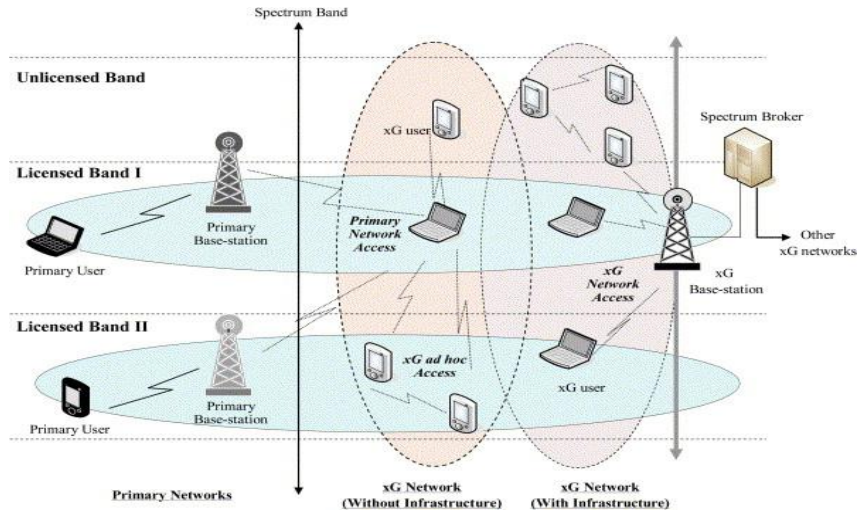

*Figure 3. Architecture des réseaux de radio cognitive. (cf Akyildiz, 2006)*

Selon l'architecture présentée dans (Akyildiz, 2006), le RRC est un ensemble de plusieurs types de réseaux qui coexistent sur les mêmes bandes spectrales. À cause de cette hétérogénéité, il existe différents types d'accès à ces réseaux (cf Ahmed Ben, 2011) :

**- accès au RRC** : les utilisateurs accèdent à leur station de base en utilisant les spectres licenciés ou non licenciés ;

**- accès au réseau ad-hoc de radios cognitives** : les utilisateurs peuvent communiquer entre eux à travers des connexions ad-hoc sur des spectres licenciés ou non licenciés ;

**- accès au réseau licencié** : les utilisateurs accèdent à la station de base des PUs en utilisant les spectres licenciés.

**3. Accès dynamique au spectre dans la cadre de la RC**

L'accès dynamique au spectre est considéré comme une solution appropriée pour résoudre le problème courant de pénurie de spectre. Les futures architectures de réseau de radio cognitive devraient être évolutive et assez coopérative, afin de fournir les meilleures solutions possibles. Le but final est d'augmenter l'utilisation de spectre par l'utilisation des techniques d'accès plus efficaces et les utilisateurs de radio cognitive pourront bénéficier de ces techniques pour pouvoir accéder au spectre requis quand et où nécessaire (si possible), à un coût accessible.

Généralement, l'accès au spectre peut être vu comme sous-ensemble de la gestion de ressource radio. Pour une meilleure explication de certains des techniques déjà proposés liés à l'accès au spectre, nous avons divisé ces travaux en différents domaines. Premièrement, nous présentons les solutions basées sur la théorie des jeux, des chaînes de Markov et des systèmes multi agent. Tandis que, concernant notre sujet principal, nous dépeignons les approches d'accès au spectre basées sur la théorie des enchères.

*3.1. Accès au spectre en utilisant la théorie des jeux*

Plus généralement, une partie de poker, la formation d'une équipe, où une négociation entre agents pour la prise de rendez-vous est autant de jeux différents obéissant à des règles spécifiques. Dans ces jeux, chaque participant ne peut être totalement maître de son sort ; on dit alors que tous les intervenants se trouvent en situation d'interaction stratégique. La théorie des jeux vise à étudier formellement ce type de jeux où le sort de chaque agent participant dans le jeu, dépend non seulement des décisions qu'il prend mais également des décisions prises par les autres agents intervenant dans le jeu. Dès lors, le "meilleur" choix pour un agent dépend généralement de ce que font les autres. Les agents participants à un jeu sont appelés joueurs. Ainsi un joueur est un agent qui pourrait représenter une entreprise, un robot, un consommateur, etc. (cf Chaib-draa., 2010)

La théorie des jeux fournit la base mathématique et statistique pour l'analyse des processus décisionnels interactifs en prévision de ce qui se produira quand les joueurs interagissent avec les conflits d'intérêts. Fondamentalement, un jeu se compose de trois partis principaux : un ensemble de joueurs, un ensemble d'actions, et un ensemble de préférences. Deux prétentions importantes sont faites, y compris la prétention de rationalité, dans laquelle, les joueurs choisissent les stratégies qui maximisent leurs gains individuels attendus, et prétention de connaissance commune, avec tous les joueurs ayant la connaissance d'autres joueurs joignant et laissant le jeu, leurs ensembles d'action, et définitions de leur rapport de préférence. (cf Mir, 2011)

Quand les joueurs agissent, les résultats entiers (utilité possible de chaque joueur) pour le jeu peuvent être dérivés par leur rapport de préférence. La propriété la plus célèbre et la plus connue de l'approche des théories des jeux est appelé équilibre de Nash (NE), qui est un concept de solution d'un jeu comportant deux ou plusieurs joueurs. En NE, on assume que chaque joueur sait les stratégies d'équilibre des autres joueurs, et aucun joueur n'a rien à gagner en changeant sa propre stratégie (cf Mir, 2011)

Largement, les concepts de théorie des jeux ont été intensivement employés pour des attributions de spectre dans les réseaux de radio cognitive, où les utilisateurs primaire et secondaires participant à un jeu, se comportent rationnellement pour choisir les stratégies qui maximisent leurs gains individuels. (cf Mir, 2011)

Les gains désirés peuvent être sous forme d'utilisation de spectre pour la durée respectée, le prix associé, le rapport signal sur bruit (SNR), le taux d'erreurs binaire (BER) et de retard, etc… (cf Mir, 2011)

Ainsi, afin de la simplicité et sur la base des comportements des joueurs dans un jeu, ces jeux sont généralement classés dans deux domaines : jeux coopératifs, et jeux compétitifs.

Dans le cadre des jeux coopératifs, les joueurs sont autorisés à communiquer avant de choisir leurs stratégies et de jouer le jeu. Ils peuvent convenir, mais aussi en désaccord sur une stratégie commune. (cf Gaoning., 2010)

Tous les joueurs sont préoccupés par tous les gains globaux et ils ne sont pas très inquiets de leur gain personnel. Certains travaux récents utilisent la théorie des jeux coopératifs pour réduire la puissance de transmission des utilisateurs secondaires afin d'éviter de générer des interférences avec les transmissions des utilisateurs primaires. (cf Beibei et al., 2007)

Au contraire, un jeu non coopératif est basé sur l'absence des coalitions (cf Mir, 2011) où chaque joueur agit individuellement pour optimiser son rendement, sans égard à la performance des autres joueurs (cf Gaoning., 2010) et donc toutes ses décisions sont prises de manière compétitive et égoïste. (cf Amraoui et al., 2012)

Dans la littérature existante, nous avons constaté que les concepts théoriques du jeu ont été largement utilisés pour des attributions de fréquences dans les réseaux RC, où lorsque les utilisateurs primaires et secondaires participent à un jeu, ils ont un comportement rationnel pour choisir les stratégies qui maximisent leurs propres gains. (cf Amraoui et al., 2012)

*3.2. Accès au spectre en utilisant les approches de Markov*

Dans ce qui précède, nous avons discuté de plusieurs aspects liés à l'application de la théorie des jeux dans les réseaux de RC. Toutefois, les approches de théorie des jeux ne savent pas modéliser l'interaction entre le SU et le PU pour l'accès au spectre. Cette modélisation peut être réalisée en utilisant efficacement les chaînes de Markov et peu de recherche ont été effectués. Dans (Chunsheng et al., 2010), un modèle de Markov est présenté, où chaque SU choisit au hasard son propre canal plutôt que d'échanger des messages de contrôle avec le SU voisin. (cf Mir, 2011)

Les processus d'accès au spectre ont aussi été modélisés en temps continu avec CTMC (Chaines de Markov en temps continu « continuous time Markov chain »). Un modèle relatif de CTMC est présenté dans (Beibei et al., 2007) pour capturer l'interaction entre les utilisateurs primaires et secondaires. Les deux modèles avec et sans files d'attente sont analysés et la dégradation du débit due à l'interférence des SU est compensée. Les modèles CTMC obtiennent des bonnes statistiques entre équité et efficacité. (cf Mir, 2011)

Un autre système CTMC de l'attribution du spectre optimal est présenté dans (Xiaorong et al., 2007), ce qui réduit la probabilité de terminaison forcée de la SU correspondant. Le même type de modèles CTMC avec et sans file d'attente sont analysés dans (Waqas et al., 2009 ; Zhang, 2008 ; Zhao et al. , 2007) et les paramètres tels que la probabilité de est considérablement réduit. (cf Mir, 2011)

Néanmoins, un nombre limité de travaux ont été réalisés pour accéder au spectre sans licence à l'aide de chaînes de Markov. Parmi ceux-ci, un célèbre modèle de CTMC est proposé dans (Xing et al., 2006) pour atteindre des assignations équitables de fréquences entre les SU. Cette approche vise plus spécifiquement sur l'utilisation efficace du spectre, en évitant les interférences. En utilisant le modèle CTMC, les auteurs tirent le temps d'utilisation du spectre et de la probabilité de blocage comme indicateurs de performance importants. Les résultats d'analyse et de simulation montrent que le modèle proposé permet d'obtenir un bon compromis entre efficacité et équité.

*3.3. Accès au spectre en utilisant les Systèmes Multi Agents (SMA)*

Les SMA ont été déjà appliqués pour l'allocation des ressources et les partages dans des contextes différents, via des applications et des réseaux sans fil actuels (cf Sycara Katia, 1998). Un des nouveaux apports des SMA est leur utilisation dans les réseaux de la radio cognitive. (cf Mir et al., 2009)

Un SMA est employé pour contrôler des ressources de spectre à travers plusieurs LANs sans fil (WLANs), situé dans une zone géographique (cf Benmammar et Krief. , 2003 ; Jrad et al. , 2005). Chaque point d'accès (AP) situé dans un WLAN contient un agent qui interagit avec les agents voisins d'AP (situés dans l'autre WLANs) pour former un SMA. (cf Jiang et al. , 2007).

L'architecture interne d'un agent se compose de deux parties : l'estimation de paramètres prédictifs, qui génère des estimations de paramètres en utilisant les caractéristiques du signal reçu à partir de l'environnement WLAN, et de l'optimisation de gestion des ressources, qui décide les bandes appropriées de spectre à choisir. L'approche proposée est expliquée conceptuellement, mais rien d'analyse et des expériences ne sont montrées.

Un schéma d'allocation distribuée et dynamique de facturation, tarification et des ressources sont examinés. Le protocole utilisé pour l'allocation des ressources radio entre les SUs et les opérateurs est nommé comme protocole d'enchères à offre scellée multi-unité qui est basé sur le concept de soumissions et d'attribution des marchandises. Un bon agent d'allocation de ressource radio (RAA) qui génère les offres pour obtenir le spectre désiré représente le comportement de chaque SU. Un agent commissaire-priseur représente le comportement de l'opérateur, ayant les capacités d'annoncer des enchères (pour l'accès au spectre) et le calcul du prix relatif. (cf Tian et al., 2010). L'écosystème se compose de plusieurs parties, y compris :

- gestion de mémoire tampon QoS (qualité de service-BM), détermine les données à envoyer en premier lieu ;

- gestionnaire de profils utilisateur (UPM), identifie les besoins de QoS et de spectre préférences pour les SU ;

- stratégie d'enchères (BRI), tire son entrée de QoS-BM de faire une offre de ressources requérante. (cf Tian et al. ,2010)

Toutes les pièces ci-dessus fonctionnent ensemble pour faire un message d'entrée (ou une offre) et ce message est alors envoyé à l'agent commissaire-priseur. A la réception d'une offre, ce dernier détermine les bandes de fréquences à attribuer à des SUs. De même, dans la deuxième approche (cf Kloeck et al., 2006), les SUs sont représentés comme des agents de consommateurs et les PUs travaillent comme agents de fournisseurs pour louer leurs spectres aux agents consommateur requérant. Un agent secondaire annonce l'accès au spectre offert (au nom d'un agent de la consommation) à l'agent fournisseur voisin, qui comprend le degré requis de contrôle sur le spectre et le prix.

En retour, l'agent fournisseur correspondant répond et s'il y a un accord, le partage du spectre est démarré, sinon les exigences des deux agents aller dans un scénario concurrentiel et l'agent avec meilleure offre remporte l'affaire. Les auteurs de (Tonmukayakul et Weiss, 2005) se concentrent sur la conception d'un algorithme centralisé de partage entre les agents de l'ISP à l'aide de ventes aux enchères.

Le commissaire-priseur centralisé attribue le spectre d'une manière libre de l'interférence, tout en maximisant la protection sociale (ou l'utilité). Les auteurs montrent comment leur régime peut réduire les frais généraux de calcul par l'allocation du spectre en une seule itération. Ils ont en outre concevoir un système de paiement qui permet aux agents d'être honnête, en leur imposant une valeur minimum fixe d'une offre. Un autre modèle d'agent basé sur l'échange ou l'allocation de spectre est présenté dans (Gopinathan et Zongpeng., 2010), où les agents acquièrent les rôles des courtiers de spectre. Le bénéfice des agents, leur probabilité d'accès au spectre et l'incertitude de la demande sont obtenus en termes de prix et de bande passante et une stratégie optimale unique (cf Gopinathan et Zongpeng., 2010). Toutefois, dans les approches concurrentielles, les agents peuvent parfois se trouver en envoyant de fausses offres, obligeant les chercheurs à développer des stratégies complexes de punir qui devraient être à l'abri de mensonge.

Les auteurs de (Ba-Lam et al., 2010) soutiennent que par le déploiement d'agents sur les SUs et en leur permettant d'apprendre par une interaction peuventt améliorer l'utilisation du spectre global et les SUs peuvent quitter facilement le spectre lorsque les PUs le récupère. Une approche d'apprentissage est proposée dans (Zhang et al., 2012), où chaque SU perçoit ses paramètres actuels de transmission et prend les mesures nécessaires lorsque le PU apparaît. Plusieurs valeurs de pénalité sont appliqués lorsque les agents SUs tentent d'interférer les uns avec les autres ou avec les PUs. Les résultats des simulations montrent que l'approche proposée améliore la transmission du SUs et les capacités de commutation de canaux par rapport à l'approche gloutonne.

En bref, les SMA peuvent être utilisés de diverses façons pour l'accès au spectre dans les réseaux de RC, mais encore le déploiement de SMA par ces infrastructures est relativement un nouveau concept.

*3.4 Accès au spectre en utilisant les Enchères*

La théorie des enchères a récemment été appliquée pour résoudre divers problèmes en ingénierie. Un processus d'enchères fonctionne comme suit. Tout d'abord, les acheteurs présentent des soumissions pour l'achat des produits, et les vendeurs soumettent les demandes de vendre des marchandises. Un produit d'enchère étant commercé peut être un article réel ou un service. Chaque offre/demande contient des informations qui indique les préférences de l'acheteur/vendeur, les exigences et les demandes pour les produits d'être commercé. Généralement, une offre doit exprimer au moins d'un prix de soumission offert et d'une quantité de produit à acheter. De même, une demande devrait contenir un prix de demande et une quantité de produit à vendre. Le prix inclus dans une offre/demande ne pourrait pas être la vraie évaluation de l'acheteur/vendeur, qui est une valeur privée évaluée par ces derniers, sur le produit pour être acheté/vendu. Il y a un agent intermédiaire désigné sous le nom du commissaire-priseur qui conduit l'enchère et dégage le marché en assortissant des offres et des demandes. Ensuite, chaque paire assortie d'acheteur-vendeur doit s'occuper à un prix d'équilibre (ou notamment au prix s'occupant). Dans quelques enchères, il y a seulement un vendeur et ce vendeur peut exécuter un rôle d'un commissaire-priseur. En conséquence, dans ce cas, les termes "commissaire-priseur" et "vendeur" peuvent être employés l'un pour l'autre. (cf Waqas et al., 2009)

Dans le contexte de la radio cognitive, les acheteurs sont des utilisateurs secondaires qui peuvent utiliser le spectre en le payant, et les vendeurs (par exemple, les utilisateurs primaires) fournissent les ressources radio aux acheteurs et recevoir des gains de ces derniers en vendant des ressources radio inutilisés ou sous-utilisés. Les produits sont des ressources radio, par exemple, la bande passante, les time-slots et les droits de transmission des données sur le réseau. En outre, les commissaires-priseurs peuvent facturer des frais de commission organisé des enchères pour les utilisateurs primaires et secondaires au revenu de gain. Grâce à la vente aux enchères, les utilisateurs primaires, les utilisateurs secondaires et commissaires-priseurs ont des motivations à participer pour obtenir des avantages dans un système de radio cognitive. (cf Vijay, 2009 ; Parsons et al., 2011 ; Zhang et al., 2012)

*3.4.1 Composants d'un marché aux enchères*

Les composants d'un marché aux enchères sont les suivantes (cf Jarraya., 2006) :

- *le vendeur* est une entité du marché qui veut vendre le produit. Un vendeur propose le prix et la quantité de produit à être négociées aux enchères.

- *l'acheteur* est une entité qui veut acheter le produit. Un acheteur soumet une offre en termes de prix et de quantité d'enchères pour acheter à travers la vente aux enchères.

- *le prix d'échange / de compensation* est le prix de chaque produit à être négociés sur un marché aux enchères. Le prix de négociation doit satisfaire le prix demandé et le prix de l'offre par le vendeur et l'acheteur, respectivement.

*3.4.2 Types plus classiques d'enchères*

Dans ***l'enchère anglaise***, le commissaire-priseur initie l'enchère en annonçant un premier prix initialement inférieur à la valeur estimée du produit. Après chaque annonce, le commissaire-priseur attend pendant un certain temps pour voir s'il existe des participants acceptant cette offre. Dès qu'un participant se manifeste, le commissaire-priseur annonce une nouvelle proposition de prix égale au dernier prix incrémenté d'une valeur appelée pas d'incrémentation. (cf Jarraya., 2006)

L'enchère se termine quand aucun des participants ne se manifeste. A ce moment-là, le commissaire-priseur compare le prix de la dernière offre acceptée avec la valeur estimée du produit. Si le prix de l'offre est supérieur à la valeur estimée alors l'enchère est accordée au participant qui s'est prononcé pour cette offre. Dans le cas contraire, les participants sont informés de l'annulation de la vente. Le manager peut réexaminer au cours de l'enchère la valeur estimée du produit, pour éviter toute annulation de la vente. (cf Jarraya., 2006)

Dans ***l'enchère hollandaise***, le commissaire-priseur annonce au départ une valeur très élevée par rapport à la valeur estimée du produit. Puis il commence par baisser le prix jusqu'à ce qu'un participant signale son acceptation du prix proposé. Le vendeur possède une estimation de la valeur du produit (prix de réserve) au-dessous de laquelle il ne vend pas le produit. Si l'enchère atteint le prix de réserve sans qu'il y ait d'acheteurs, l'enchère se termine. (cf Jarraya., 2006)

Dans ***l'enchère Vikrey*** (Enveloppe scellée au second prix), les propositions sont privées, aucun participant ne connaît le contenu des autres enchères. Les enchères Vickrey permettent de vendre un article unique, comme les enchères anglaises. La différence est que le soumissionnaire le plus élevé obtient l'article au prix offert par le deuxième plus haut soumissionnaire. (cf Jarraya., 2006)

Dans ***l'enchère au premier prix***, des enchères sont soumises au vendeur sous un certain format, sans qu'aucun des participants ne sache le contenu des autres enchères. Le gagnant paye le prix qu'il avait soumis. (cf Jarraya., 2006)

Dans ***le double enchère***, les participants sont des acheteurs et des vendeurs qui négocient sur un même produit. L'enchère peut démarrer par l'envoie d'une offre. (cf Jarraya., 2006)

Les vendeurs envoient des offres qui démarrent avec le plus haut prix puis elles décroissent, et des propositions sont faites par les acheteurs dont les prix évoluent dans le sens inverse des offres. Les offres et les propositions sont publiques, chaque participant est au courant de toutes les offres et les propositions soumises (l'émetteur et le prix proposé). La négociation peut s'achever si un acheteur accepte une offre proposée par un vendeur ou si un vendeur accepte une proposition d'un acheteur. (cf Jarraya., 2006)

*3.4.3 Quelques travaux relatifs*

Dans (Zhu et al., 2009), les auteurs ont étudié le problème d'accès au spectre dans le cadre de la radio cognitive unique et multicanal. Un cadre fondé sur l'enchère répétée a été adoptée. Dans l'accès au spectre à canal unique, les SUs participent de manière sélective dans la vente aux enchères en fonction de leur évaluation et historique des enchères passé. Dans les réseaux multicanal, une approche non-regret a été proposée. Sa performance a été montrée pour être nettement mieux qu'une solution gourmande naïve.

Dans les solutions basées sur les enchères, chaque canal est assigné à un seul réseau, c'est à dire qu'il n'y a pas la notion de SU et de PU dans le même canal. Dans la littérature, deux possibilités s'offrent (cf Amraoui et al., 2013) :

- soit le régulateur alloue les canaux aux utilisateurs primaires, ces derniers allouent indépendamment les portions inutilisées de leur canal aux SU ;

- soit le régulateur alloue le droit d'être SU ou PU dans le canal.

Dans (Chen et al., 2008), les auteurs adoptent les mécanismes d'enchères au second prix pour résoudre le problème d'allocation du spectre ainsi que les problèmes de détection des bandes libres à la radio cognitive. Ils introduisent la notion d'argent fictif en réalisant le système de payement en temps réel.

Dans une autre approche, les auteurs proposent une nouvelle vente aux enchères basées sur le temps d'attente pour l'allocation dynamique du spectre. Avec cette approche, les soumissions des utilisateurs sont leur temps d'attente au lieu d'argent. (cf Guangen et al., 2011)

Dans (Lin et al., 2010), les auteurs proposent un cadre d'enchères pour les réseaux de radio cognitive pour permettre aux SUs non autorisés de partager le spectre disponible avec des PUs autorisés de manière équitable et efficace, soumis à la contrainte de la température de l'interférence à chaque PU. Ils étendent ensuite le cadre de la vente aux enchères proposée pour le scénario le plus difficile avec des bandes de fréquences libres où le SU aura le choix entre payer et avoir une bonne QoS ou bien accéder aux canaux gratuitement et risquer de rencontrer des interférences avec les utilisateurs.

Dans certains scénarios, le spectre est utilisé de manière plus efficace en accordant une bande à plusieurs SUs simultanément, ce qui le distingue des enchères traditionnelles où un seul utilisateur peut être le gagnant. Toutefois, la vente aux enchères multi-gagnantes est un nouveau concept qui pose de nouveaux défis aux mécanismes d'enchères traditionnelles. Par conséquent, dans (Yongle et al., 2008), les auteurs proposent un mécanisme efficace de la vente aux enchères du spectre multi-gagnant résistant à la collusion où plusieurs SU peuvent gagner l'accès à un seul canal.

**4. Implémentation de la solution proposée**

Dans ce papier, nous avons choisis un type particulier d'enchères, c'est l'enchère à enveloppe scellée, nous l'avons implémenté avec et sans programmation dynamique et comparé les résultats obtenus avec ceux de la méthode FIFO'[1] dans le but de mesurer l'importance de l'enchère pour le partage du spectre mais aussi l'importance de combiner la programmation dynamique avec les enchères.

Notre approche est basée sur les SMA et implémentée à l'aide de la plateforme JADE qui facilite la méthodologie SMA car c'est un langage de programmation d'agents qui permet de suivre les traces des messages échangés et afficher chaque message dans la fenêtre du sniffer. Nos agents (PU et SUs) utilisent le type de négociation « plusieurs à un c.-à-d. plusieurs SUs et un seul PU».

*4.1. Topologie du réseau utilisé*

Dans la littérature, la plupart des inconvénients et des problèmes liés aux enchères sont en rapport avec le régulateur (initiateur), ce dernier peut avoir un comportement mensonger, il peut aussi utiliser de faux participants pour faire augmenter l'évaluation de l'objet. (cf Amraoui et al., 2013)

Pour éviter ce type de problème, nous utilisons dans ce travail une architecture de système multi-agents coopératifs pour la gestion des ressources radio, c'est une architecture de réseau sans infrastructure « réseau Ad-hoc ». Nous considérons que tous nos agents sont fixes durant leurs mises en œuvre, car ils travaillent localement, chacun dans son propre site et communiquent entre elles de façon directe.

Autrement dit, la communication se fera directement entre le PU et les SUs (il n'y a pas la notion de commissaire-priseur c.-à-d. l'enchère se fait directement entre le vendeur et l'acheteur). La Figure 4 illustre la topologie du réseau que nous avons utilisé :

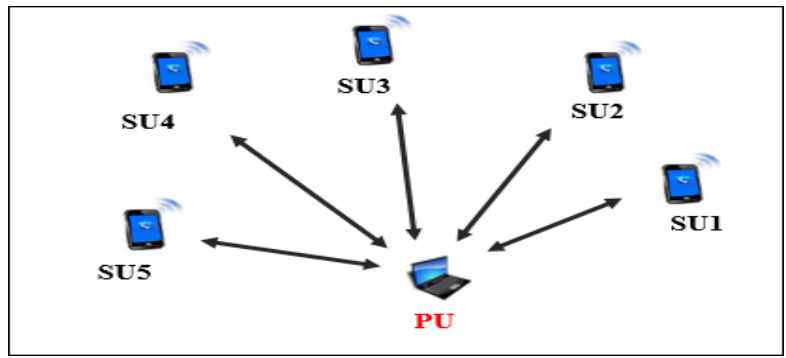

*Figure 4. Topologie du réseau (mode Ad-hoc).¶¶*

---

[1]FIFO' : FIFO mais sans blocage des demandes insatisfaites.

*4.2. Scénario proposé*

Nous considérons un scénario où le PU et les SUs sont connectés en mode « Ad-hoc » avec le type particulier de négociation «plusieurs à un» tels que l'agent PU négocie avec les agents SUs qui ont besoin de bandes libres pour maximiser l'accès dynamique au spectre.

Dans notre situation, le PU et les SUs négocient leur accord sur la base de certains critères tels que le prix et le nombre de canaux. La Figure 5 illustre le scénario qui sera traité dans la suite de ce travail :

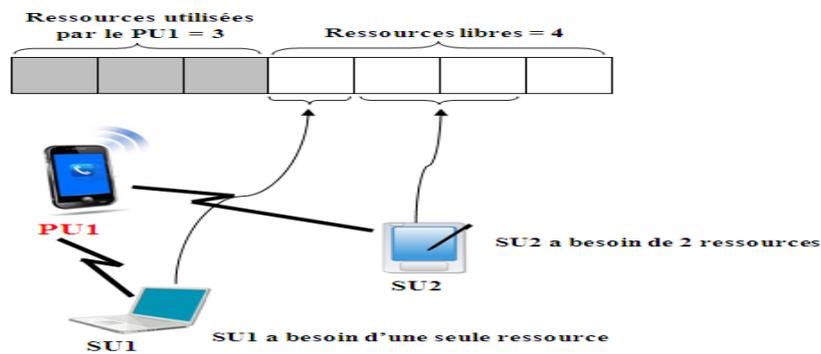

*Figure 5. Scénario proposé.*

*4.3. Algorithme proposé*

Les auteurs dans (Amraoui et al., 2013), ont comparé les enchères à un tour avec les enchères à plusieurs tours (nommées aussi enchères classiques) mais en les combinant systématiquement avec la programmation dynamique, et aussi sans passer par une vraie simulation du comportement de négociation entre agents. C'est-à-dire en ignorant les interactions entre SUs et PU. Le travail des auteurs est évalué uniquement en centralisé coté PU.

Dans cet article, nous avons opté pour une comparaison entre les enchères à un seul tour (enveloppe scellée au premier prix) et les enchères à un seul tour avec programmation dynamique mais dans le cadre d'un vrai comportement de négociation à l'aide d'une simulation avec l'outil Jade. L'objectif est de montrer l'apport de la programmation dynamique pour les enchères. Ensuite nous avons comparé les résultats obtenus avec ceux de la méthode FIFO'.

Dans ce qui suit, nous allons noter :

- **nb** : le nombre de SU.

- **m** : le nombre de canaux libres coté PU.

- **W** : tableau de taille égale au nombre de SU, **W[i]** est le nombre de canaux demandés par SUi.

- **C** : tableau de taille égale au nombre de SU, **C[i]** représente le prix proposé pour **W[i]** par SUi.

La fonction monotone croissante à optimiser est :

$$Max \sum_{i=0}^{nb-1} C[i] \qquad (1)$$

La contrainte est :

$$\sum_{i=0}^{nb-1} W[i] \le m \qquad (2)$$

L'algorithme proposé dans (Amraoui et al., 2013) pour la programmation dynamique est le suivant :

*Algorithme 1. Algorithme de programmation dynamique adapté dans le cadre de la RC*

```
1 : fonction COUT (W, C, m)
2 : n ← longueur (C)
3 : tab : tableaux à 2 dimensions de taille (n+1)*(m+1)
4 : pour j=0 à m faire
5 :     tab[0][j] ← 0
6 : fin boucle
7 : pour i=1 à n faire
8 :     pour j=0 à m faire
9 :         si (j < W[i-1]) alors
10:             tab[i][j]=tab[i-1][j]
11:         sinon
12:             tab[i][j]=Max(tab[i-1][j],C[i-1]+tab[i-1][j-W[i-1]])
13:         fin condition
14:     fin boucle
15: fin boucle
16: retourner tab[n][m]
17: fin fonction
```

## 4.4. Etude comparative

Pour la simulation, nous avons utilisé le même jeu de données pour les trois méthodes (enchère à enveloppe scellée avec et sans programmation dynamique et FIFO') pour pouvoir comparer les résultats obtenus. Le jeu de données utilisé est : nb=5 et m=4, C = {300, 354.35, 212.6, 141.7, 141.68} et W = {6, 5, 3, 2, 2}.

Nous avons implémenté l'algorithme de l'enchère à enveloppe scellée avec et sans programmation dynamique coté PU et ensuite comparé les résultats obtenus avec celui de la méthode FIFO' sachant que nous supposons qu'il y a 5 SUs au même endroit.

*4.4 .1 Comparaison en termes d'efficacité*

Ici, nous parlons de nombre de SU satisfaits. Pour cela, nous avons comparé les trois algorithmes et nous remarquons les résultats obtenus dans la Figure 6 qui montre l'impact des enchères sur le nombre de SUs satisfaits.

Le jeu de données utilisé est : C = {300, 354.35, 212.6, 141.7, 141.68} et W = {6, 5, 3, 2, 2} et nous fixons le nombre de SUs à 5 (nb=5) et nous varions le nombre de canaux (m = 1…10).

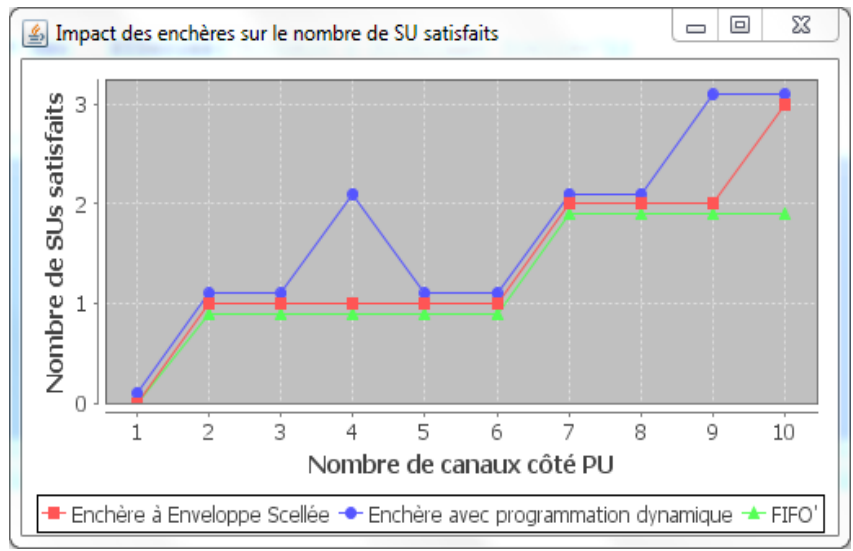

*Figure 6. Impact des enchères sur le nombre de SUs satisfaits.*

D'après la Figure 6, nous remarquons que le nombre de SUs satisfaits avec l'enchère en utilisant la programmation dynamique est supérieur par rapport à celui obtenu avec FIFO' et l'enchère à enveloppe scellée.

*4.4.2. Comparaison en termes de gain de PU*

Dans cette section, nous allons montrer l'impact des enchères sur le gain obtenu par le PU, pour cela nous allons faire une comparaison entre l'enchère avec et sans programmation dynamique et avec la technique FIFO'.

Les résultats obtenus sont illustré dans la Figure 7.

Le jeu de données utilisé est : nb=5, C = {286, 209, 141, 489, 105} et W = {3, 2, 1, 3,2} avec un « m » qui varie entre 1et 10.

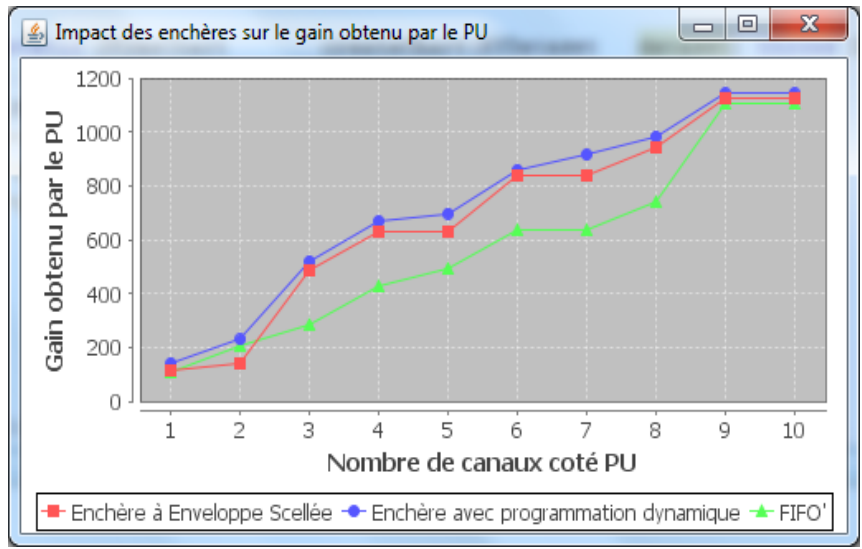

*Figure 7. Impact des enchères sur le gain obtenu par le PU.*

Nous remarquons que l'utilisation des enchères avec la programmation dynamique est plus bénéfique pour le PU car ses gains sont beaucoup plus importants par rapport à l'utilisation des enchères à enveloppe scellée ou l'utilisation de la technique FIFO'.

*4.4.3. Comparaison en termes de temps requis*

Pour connaître le temps (en milli seconde « ms ») du traitement coté PU pour qu'un nombre de SUs peut être satisfaits, nous avons fait une comparaison entre l'utilisation des deux types d'enchères avec la méthode FIFO'.

Pour établir cette comparaison, nous considérons le cas d'un nombre variable de SUs (nb = 1…10), et d'un nombre fixe de canaux (m=5).

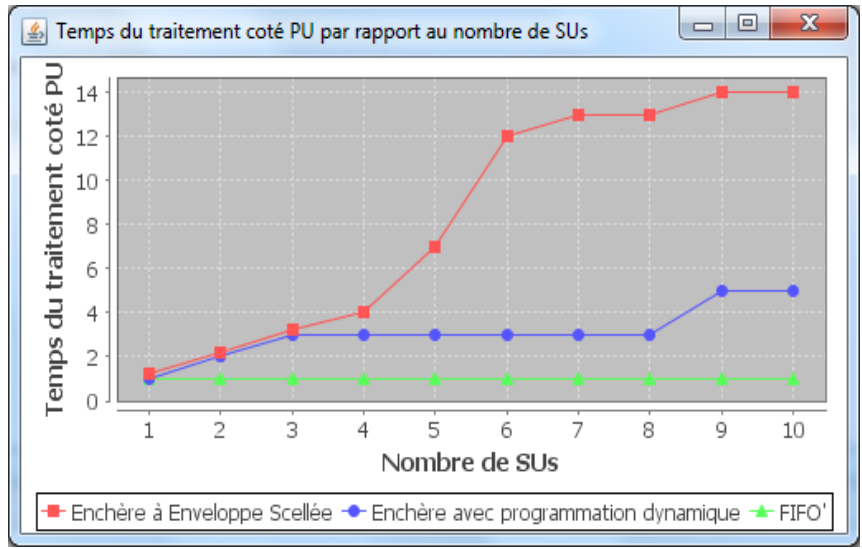

*Figure 8. Temps du traitement coté PU par rapport au nombre de SUs.*

Comme le montre la Figure 8, avec les deux types d'enchères, le temps du traitement côté PU s'augmente en fonction de l'augmentation du nombre de SUs. C'est clair car avec plus de SUs disponibles, le PU prend plus de temps pour choisir les meilleures offres proposé par ces SUs. Mais l'utilisation des enchères avec programmation dynamique reste mieux que l'enchère à enveloppe scellée en termes de temps requis car l'enchère à enveloppe scellée dans notre cas a besoin de beaucoup de temps pour trier les demandes reçu selon le prix unitaire de chaque SU. Par contre l'idée de l'enchère avec programmation dynamique est de classer les SUs selon le prix proposé pour la totalité des canaux pour obtenir une solution optimale. Dans le cadre général, la programmation dynamique prend beaucoup de temps pour trier ces objets mais dans notre cas et puisque la taille du tableau est raisonable, elle nous donne un résultat meilleur que l'enchère à enveloppe scellée (la taille du tableau, dans ce cas est le nombre de SUs * le nombre de canaux). Pour la technique FIFO', le temps requis est toujours égal à 1ms car le traitement est limité (requête/réponse) le premier entré est le premier servi sans bloquer les demandes.

*4.4.4 Comparaison en termes de temps de réponse coté SU*

La Figure 9 montre l'impact du taux d'arrivée des SUs sur le temps de réponse toujours coté SU, pour se faire, nous varions le taux d'arrivé des SUs de 1s à 10s, le PU est lancé après 1s du lancement de SU10. Ce graphe montre le temps de réponse uniquement pour l'enchère (à enveloppe scellée avec et sans programmation dynamique), car pour FIFO' c'est requête/réponse donc le traitement est limité.

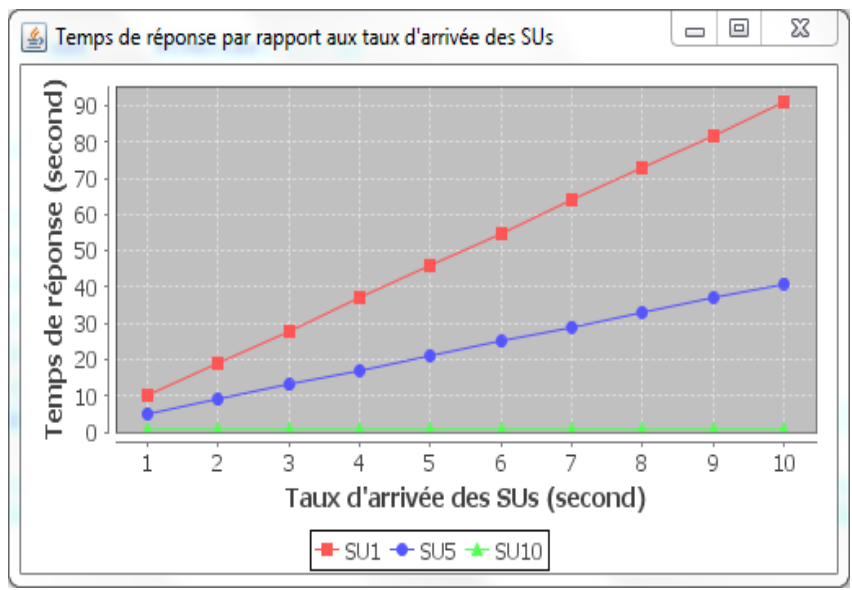

*Figure 9. Temps de réponse par rapport aux taux d'arrivée des SUs.*

Il faut remarquer ici que le temps de réponse coté SU est le temps attendu pour avoir une réponse de la part du PU, car pour les enchères il faut attendre jusqu'à l'arrivée du dernier SU et le lancement du PU. (Le coût du traitement en ms coté PU est toujours négligeable).

Ce graphe montre que le temps de réponse pour un SU donné s'augmente en fonction du taux d'arrivée des SUs sauf dans le cas du dernier SU arrivée qui va toujours attendre une seconde (le temps pour lancer le PU). Nous remarquons aussi que plus l'arrivé d'un SU est proche de celui du dernier SU, plus son temps de réponse est diminué. Et plus l'arrivé d'un SU est loin de celui du dernier SU arrivée plus le temps de réponse est augmenté. Pour être plus précis, si X est le taux d'arrivée des SUs :

- *SU1* va attendre 9*X+1 car 9 SUs arriveront derrière lui chaque X seconds, le PU est lancé après SU10 de 1s (courbe rouge).

- *SU5* va attendre 5*X+1 car 5 SUs arriveront derrière lui chaque X seconds, le PU est lancé après SU10 de 1s (courbe bleu).

- Par contre, *SU10* arrive le dernier donc il va toujours attendre 1s, le temps pour lancer le PU (courbe verte).

**5. Conclusion**

Dans ce papier, nous avons combiné la théorie des enchères avec les systèmes multi agents pour résoudre le problème de l'encombrement du spectre. Pour cela, nous avons donné les notions les plus importantes concernant la topologie et les outils utilisés pour faire simuler l'accès dynamique aux spectres en utilisant un type particulier d'enchères qui est l'enchère à enveloppe scellée au premier prix.

Les résultats obtenus montrent que quel que soit le nombre de canaux demandés, l'utilisation des enchères avec la programmation dynamique est mieux que les enchères à enveloppe scellée et FIFO' car la procédure est plus rapide et a beaucoup d'avantages en termes de nombre de SUs satisfaits, gain obtenu par le PU, temps de traitement côté PU et finalement temps de réponse coté SU.

Comme conclusion nous pouvons dire que notre approche a prouvé qu'il est préférable d'utiliser les enchères avec la programmation dynamique pour résoudre le problème de l'encombrement du spectre et pour une gestion plus efficace de celui-ci car tous les utilisateurs sont gagnants. Les SUs gagnent en satisfaisant leurs besoins et le PU gagne en maximisant son gain.

*Biographie*

Badr Benmammar est un enseignant chercheur à UABT et au laboratoire de Télécommunication de Tlemcen (LTT), ses activités de recherches portent sur les réseaux de radio cognitive et sur la gestion de la qualité de service et de la mobilité dans les réseaux sans fil, il est l'auteur du livre « Radio Resource Allocation and Dynamic Spectrum Access » apparu chez Wiley-ISTE en 2013, ainsi que de nombreuses revues scientifiques dans le domaine des réseaux.